\documentclass[10pt,preprintnumbers,superscriptaddress,amsmath,amssymb,aps,showpacs, twocolumn]{revtex4-2}

\usepackage{stmaryrd}
\usepackage{amsmath}
\usepackage{amssymb}
\usepackage{float}
\usepackage{CJK}
\usepackage{array, color}
\usepackage{bm}
\usepackage{tabularx}
\usepackage{multirow}
\usepackage{threeparttable}
\usepackage{titletoc}
\usepackage{booktabs}
\usepackage{ulem}
\usepackage[colorlinks,linkcolor=blue,anchorcolor=blue,citecolor=blue,urlcolor=blue,driverfallback=dvipdfm]{hyperref}

\usepackage{graphicx}
\usepackage{subfigure}
\usepackage{latexsym,bm}

\date{\today}
\begin{document}
\draft


\title{Anomalous thermal transport across the superionic transition in ice}


\author{Rong~Qiu}
\author{Qiyu~Zeng}
  \affiliation{Department of Physics, National University of Defense Technology, Changsha 410073, China}
  \affiliation{Hunan Key Laboratory of Extreme Matter and Applications, National University of Defense Technology, Changsha 410073, China}

\author{Han~Wang}
  \affiliation{Laboratory of Computational Physics, Institute of Applied Physics and Computational Mathematics, Beijing 100088, China}

\author{Dongdong~Kang}
  \affiliation{Department of Physics, National University of Defense Technology, Changsha 410073, China}
  \affiliation{Hunan Key Laboratory of Extreme Matter and Applications, National University of Defense Technology, Changsha 410073, China}

\author{Xiaoxiang~Yu}
  \email[Corresponding author: ]{xxyu@nudt.edu.cn}
\author{Jiayu~Dai}
  \email[Corresponding author: ]{jydai@nudt.edu.cn}
  \affiliation{Department of Physics, National University of Defense Technology, Changsha 410073, China}
  \affiliation{Hunan Key Laboratory of Extreme Matter and Applications, National University of Defense Technology, Changsha 410073, China}

\begin{abstract}

Superionic ices with highly mobile protons within the stable oxygen sub-lattice occupy an important proportion of the phase diagram of ice and widely exist in the interior of icy giants and throughout the universe. Understanding the thermal transport in superionic ice is vital for the thermal evolution of icy planets. However, it is highly challenging due to the extreme thermodynamic conditions and dynamical nature of protons, beyond the capability of the traditional lattice dynamics and empirical potential molecular dynamics approaches. In this work, by utilizing the deep potential molecular dynamics approach, we investigate the thermal conductivity of ice-VII and superionic ice-VII" along the isobar of $p = 30\ \rm{GPa}$. A non-monotonic trend of thermal conductivity with elevated temperature is observed. Through heat flux decomposition and trajectory-based spectra analysis, we show that the thermally-activated proton diffusion in ice-VII and superionic ice-VII" contribute significantly to heat convection, while the broadening in vibrational energy peaks and significant softening of transverse acoustic branches lead to a reduction in heat conduction. The competition between proton diffusion and phonon scattering results in anomalous thermal transport across the superionic transition in ice. This work unravels the important role of proton diffusion in the thermal transport of high-pressure ice. Our approach provides new insights into modeling the thermal transport and atomistic dynamics in superionic materials.

\end{abstract}
\maketitle



As one of the most abundant substances in Earth and the universe, ice is of vital importance from a scientific perspective and attracts wide research interests. Especially, in the interior conditions of icy moons, where pressure ranges from 2 GPa to hundreds of GPa and temperature ranges from 300 K to 4000 K, high-pressure ice phases (VII/VII"/X) is expected to widely exist \cite{ref:prakapenka2021}. 
These phases present the same body-centered cubic (BCC) oxygen sub-lattice but differ in the dynamics of hydrogen atoms (protons)\cite{kangdd}. For the molecular crystal ice-VII, the orientation of hydrogen-bonding is disordered and continually changing as in hexagonal ice, obeying the 'ice rule' \cite{ice_rule}.
When temperature grows above thousands of Kelvin, ice-VII transforms into the superionic phase VII" \cite{ref:cavazzoni1999}. It has been suggested that the suitable conditions for superionic ice lie deep inside the watery giants Uranus and Neptune and may be common throughout the Universe \cite{water_book, ref:prakapenka2021}. VII" is characterized by highly mobile hydrogen ions (protons), behaving like a liquid and moving within the BCC oxygen sub-lattice. The difference in the behavior of protons can result in anomalies in thermodynamic and transport properties of ice.

The occurrence and geodynamic behaviors of these high-pressure ice polymorphs (MPa-GPa range) have important effects on the thermal evolution of icy planets. On this issue, thermal conductivity serves a key role for in-depth understanding. However, despite the enormous phase transition regime and proton transfer dynamics explored by previous efforts \cite{ref:schwegler2008, ref:Hernandez2016, ref:hernandez2018, ref:Queyroux2020, ref:prakapenka2021}, it seems likely that we still know little about the thermal transport properties of dense ice across superionic transition.

Existing experimental efforts had been pursued to measure the thermal conductivity of ice-VII up to 20 GPa \cite{ref:andersson2005, ref:chen2011}, but still far from the condition of superionic regime due to the limitation of experimental techniques under extreme conditions. From a theoretical point of view, the dynamical nature of protons prevents the most commonly used tool, the lattice dynamics approach, from tackling these issue. Another way to obtain thermal conductivity is molecular dynamics simulation. However, \textit{ab initio} method requires an expensive computational cost to reach the long-time trajectories with large simulation size required for estimation of correlation function \cite{ref:grasselli2020}. Moreover, the diverse local environments that characterize the different relevant phases of water make classical force fields unfit for an accurate simulation of their properties. 
Until now, the microscopic mechanism determining the thermal conductivity of superionic ice at high pressure remains unclear.




\begin{figure*}[htbp]
\centering
  \includegraphics[width=0.9\textwidth]{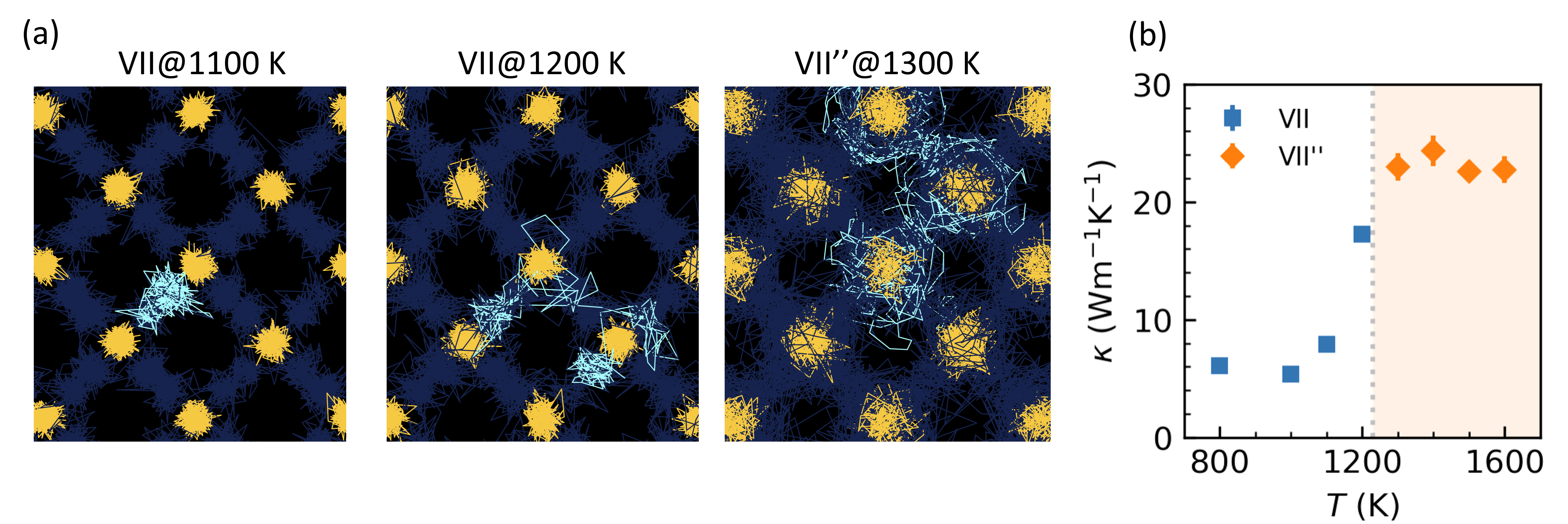}
  \caption{ (a) Atomic trajectories of ice-VII and VII" at different temperatures  during a 20-ps long run. The oxygen and hydrogen atoms are orange and blue respectively. The cyan color is used to highlight the selected hydrogen atoms that undergo transitions from bonded states between two adjacent oxygen atoms to superionic states around different oxygen atoms. (b) Temperature-dependent thermal conductivity $\kappa$ of ice-VII and VII" along the isobar of P = 30 GPa. The gray vertical dashed line denotes the VII-VII" phase boundary obtained from previous work \cite{ref:zhang2021phase}. }
  \label{fig:1}
\end{figure*}

Recent advances in machine-learning potential surface allow a full quantum-mechanical, \textit{ab initio} treatment of the interatomic interactions efficiently. The deep potential water model is reported to predict a phase diagram close to experiments \cite{ref:zhang2021phase}, and its following applications have demonstrated its success in estimating the thermal conductivity of water at extreme conditions \cite{ref:zhang2023thermal,ref:Yang_2022}. 
Therefore, in this work, we adopt the DP-SCAN water model and conduct a series of deep potential molecular dynamics (DPMD) simulations to obtain the thermal conductivity of VII and VII", as well as diffusion coefficient, spectral energy density (SED), and dynamic structure factor (DSF), to understand the heat transport and to unravel the impact of mobile protons across superionic phase transition. 


\textit{Computational Details} 
The DP model was trained with the DeePMD-kit package \cite{ref:wang2018deepmd, ref:dpkitv2} using diverse ice crystal and liquid phase covering from ambient condition to extreme thermodynamic state (p = 50 GPa, T = 2400 K). The training data were obtained from density functional theory calculations using the strongly constrained and appropriated normed (SCAN) exchange-correlation functional. More details can be found in \cite{ref:zhang2021phase}.

With DPMD simulation, the lattice thermal conductivity $\kappa$ is obtained from the integration of the heat current autocorrelation function (HCACF), known as Green–Kubo formula \cite{ref:mcquarrie1965statistical}. We performed a series of DPMD simulations with the LAMMPS package \cite{ref:plimpton1995fast}. A large supercell containing 1,296 atoms is used to overcome the size effect (see Fig. S1 in the SI). The timestep was set to 0.5 fs and the Nos\'{e}-Hoover thermostat \cite{ref:nose1984unified,ref:hoover1985canonical} was employed in the NVT ensemble.
After a thermalization stage of 20 ps, the ensemble is switched into the NVE ensemble to calculate the HCACF during the next 320 ps with the correlation time set to 32 ps. To provide a representative sample for the relevant statistical analysis, each (P, T) case repeats 20 times, with independent initial velocity distribution. We note that such computation complexity can hardly be achieved by the traditional AIMD method.


\textit{Non-monotonic behavior of thermal conductivity at elevated temperature.} The isobar of $p = 30\ \rm{GPa}$ is chosen to investigate the thermal and proton transport of ice-VII and VII". As the temperature increases from 800 K to 1600 K, our DPMD simulations reproduce the superionic phase transition of ice reported in previous works \cite{ref:Hernandez2016, ref:Queyroux2020, ref:zhang2021phase}. The atomic trajectories at temperatures near the phase boundary are shown in Fig.\ref{fig:1}(a). More atomic trajectories can be seen in Fig. S2 in SI. We can easily identify the different behaviors of protons in the oxygen sub-lattice. At low temperatures, the hydrogen atoms are bonded and only vibrate within the O-H···O bonds. 
At 1200 K close to the phase boundary, the hydrogen atoms begin to initiate hopping between different O-H···O bonds but remain bonded with oxygen atoms. The proton can migrate from an O-H···O bond to another, leading to a fast change in the orientation of water molecules. At higher temperatures, the protons diffuse freely out of the bcc oxygen sub-lattice. Namely, the system transits into a superionic phase.

Correspondingly, $\kappa$ exhibits a non-monotonic trend, as shown in Fig. \ref{fig:1}(b). Firstly, $\kappa$ decreases from $6.12\pm 0.14\ \rm{W m^{-1} K^{-1}}$ at 800 K to a minimum value of $5.37\pm 0.21\ \rm{W m^{-1} K^{-1}}$ at 1000 K. Then a significant increase in $\kappa$ to a three-fold value of $17.28\pm 0.67\ \rm{W m^{-1} K^{-1}}$ at T = 1200 K is observed. As the ice transits from VII phase into VII" phase, $\kappa$ gradually increases and finally reaches a plateau value of $22.72\pm 1.13\ \rm{W m^{-1} K^{-1}}$ at 1600 K.  

\begin{figure}[htbp]
\centering
  \includegraphics[width=0.4\textwidth]{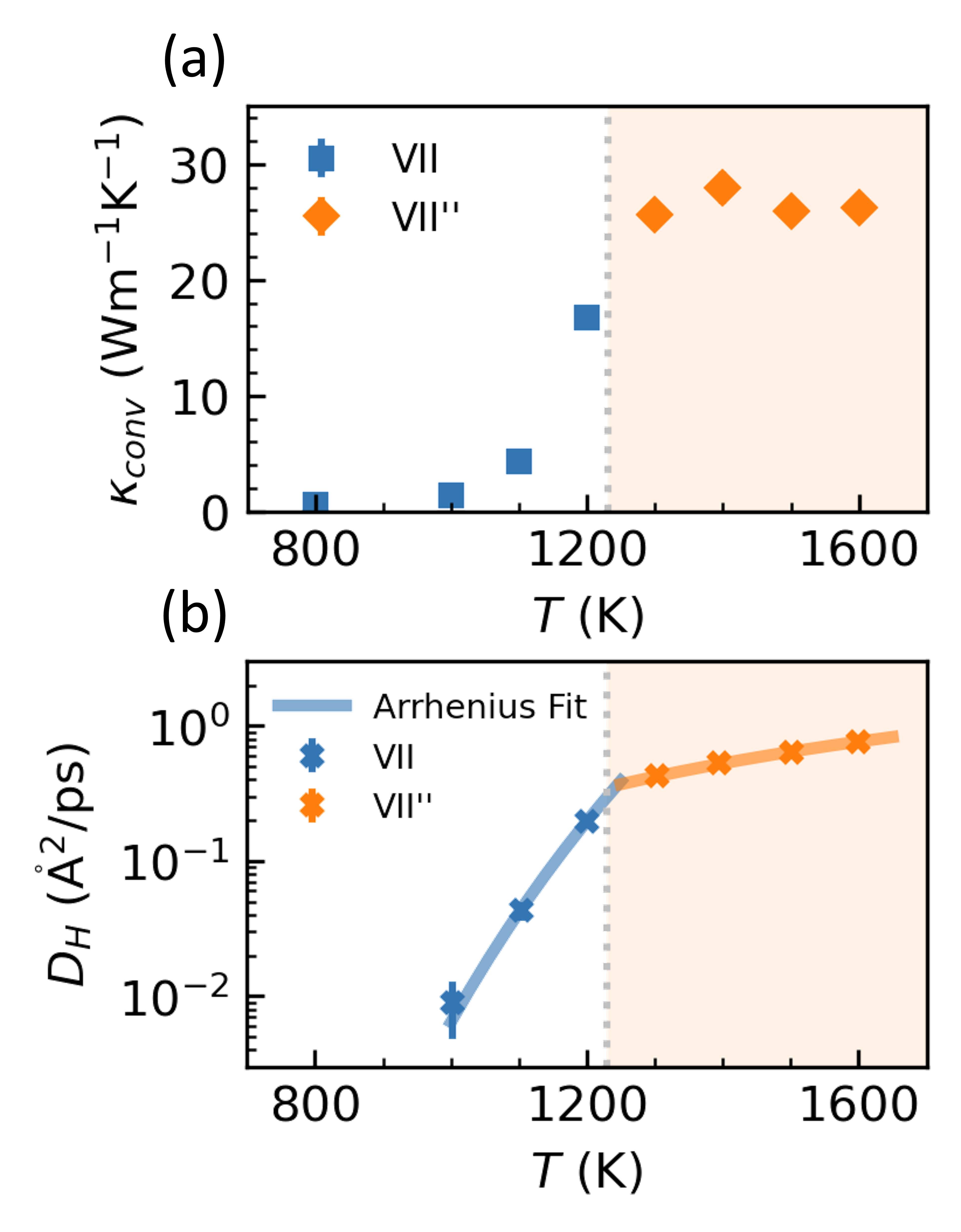}
  \caption{Temperature dependence of (a) thermal conductivity contributed by heat conduction $\kappa_{conv}$ (b) diffusion coefficient along the isobar of p = 30 GPa. Blue and orange markers denote the results of ice-VII and superionic VII", respectively. The gray vertical dashed line denotes the VII-VII" phase boundary obtained from previous work \cite{ref:zhang2021phase}.}
  \label{fig:2}
\end{figure}

\textit{Dominant role of proton diffusion in heat convection.} The anomalous non-monotonic trend of $\kappa$ is attributed to the significant change in proton transport across the superionic transition. We extract the contributions of heat convection and conduction to $\kappa$ ($\kappa_{conv}$ and $\kappa_{cond}$) by decomposing the heat current into a heat convection term and a heat conduction term, respectively (see more details in SI). As depicted in Fig. \ref{fig:2}(a), $\kappa_{conv}$ shows a monotonic increasing trend with increasing temperature. At low temperatures, $\kappa_{conv}$ is close to zero. At temperatures near the phase boundary, $\kappa_{conv}$ increases sharply. At higher temperatures, $\kappa_{conv}$ gradually converges to a plateau similar to $\kappa$. $\kappa_{conv}$ overwhelms the $\kappa$ after the superionic transition, because of the fast diffusion of protons. The behavior of $\kappa_{conv}$ dominates the dramatic increase and saturation of $\kappa$, highlighting the importance of proton transfer dynamics in understanding the heat transport in ice-VII and superionic VII".

We plot the mean square displacements (MSD) of oxygen and hydrogen atoms in Fig. S3 in SI. The diffusion coefficient can be estimated from the slope of MSD from DPMD trajectories. Oxygen atoms have a flat curve of MSD and thus a diffusion coefficient close to zero. In comparison, the MSD curves of proton show diffusion characteristics and different slopes at different temperatures. As shown in Fig. \ref{fig:2}(b), the diffusion coefficient of proton $D_H$ increases by two orders of magnitude, as the temperature increases from 1000 K to 1600 K. Besides, the diffusion behavior of proton at elevated temperatures can be well described as an Arrhenius-like process, which is given as $D(T) = A e^{-E_a/k_B T}$, where $T$ is temperature, $A$ is a prefactor, $E_a$ is the activation energy of the hopping mechanism leading to particle diffusion, and $k_B$ is Boltzmann constant. By fitting the $D_H$ with the Arrhenius model, the $E_a$ of ice-VII and VII" is $1.77\ \rm{eV}$ and $0.36\ \rm{eV}$ respectively. The much smaller $E_a$ of ice-VII" indicates the weaker confinement and stronger diffusion of protons in superionic phase. The temperature where the diffusion behavior of proton changes is consistent with the phase boundary of superionic transition.

\begin{figure}[htbp]
\centering
  \includegraphics[width=0.4\textwidth]{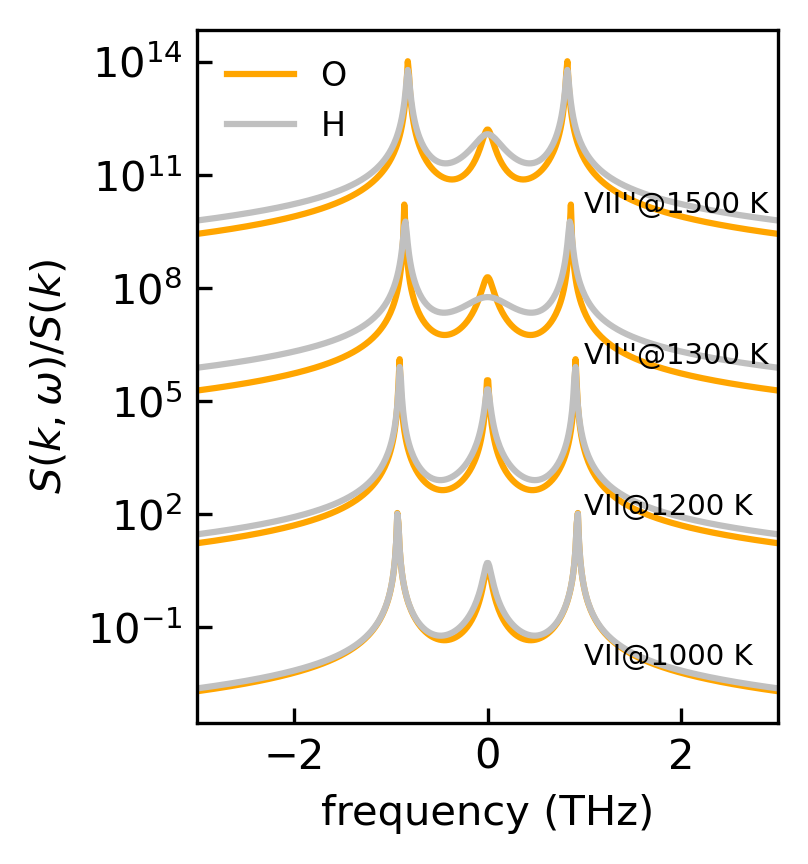}
  \caption{ Normalized dynamic structure factor $S(k,\omega)$ of hydrogen and oxygen atoms along the isobar of p = 30 GPa at $k = 0.07\ \rm{\mathbf A^{-1}}$.}
  \label{fig:3}
\end{figure}

Except for the mass diffusion, we further investigate the thermal diffusion based on the dynamic structure factor $S(k,\omega)$ (see more details in SI). The central Rayleigh peak encodes the thermal diffusion process as the wavenumber is small enough to reach the hydrodynamic regime \cite{hansen1990theory, zeng2021ab}. At the hydrodynamic limit, the shape of this central peak can be described by a Lorentzian function with peak width relating to the thermal diffusivity $D_T$, which gives $S(k,\omega) \propto 2D_Tk^2/(\omega^2+(D_Tk^2)^2)$.

We calculate the $S(k,\omega)$ of oxygen and hydrogen atoms. By fitting the DPMD results with the hydrodynamic expression, we present the normalized Rayleigh-Brillouin triplets for both H-contributed and O-contributed $S(k,\omega)$ at a small wavenumber of $k = 0.07\ \rm{\mathbf A^{-1}}$. As presented in Fig. \ref{fig:3}, as temperature increases, the broadening of central Rayleigh peaks for protons is more significant than that for oxygen atoms, indicating a more significant increase in thermal diffusion of protons compared with that of oxygen atoms. The sharp increase in $\kappa$ across the superionic transition stems from the dramatically enhanced diffusion of protons and thus sharply increased thermal diffusion of protons. 

\begin{figure*}[htbp]
\centering
  \includegraphics[width=1.0\textwidth]{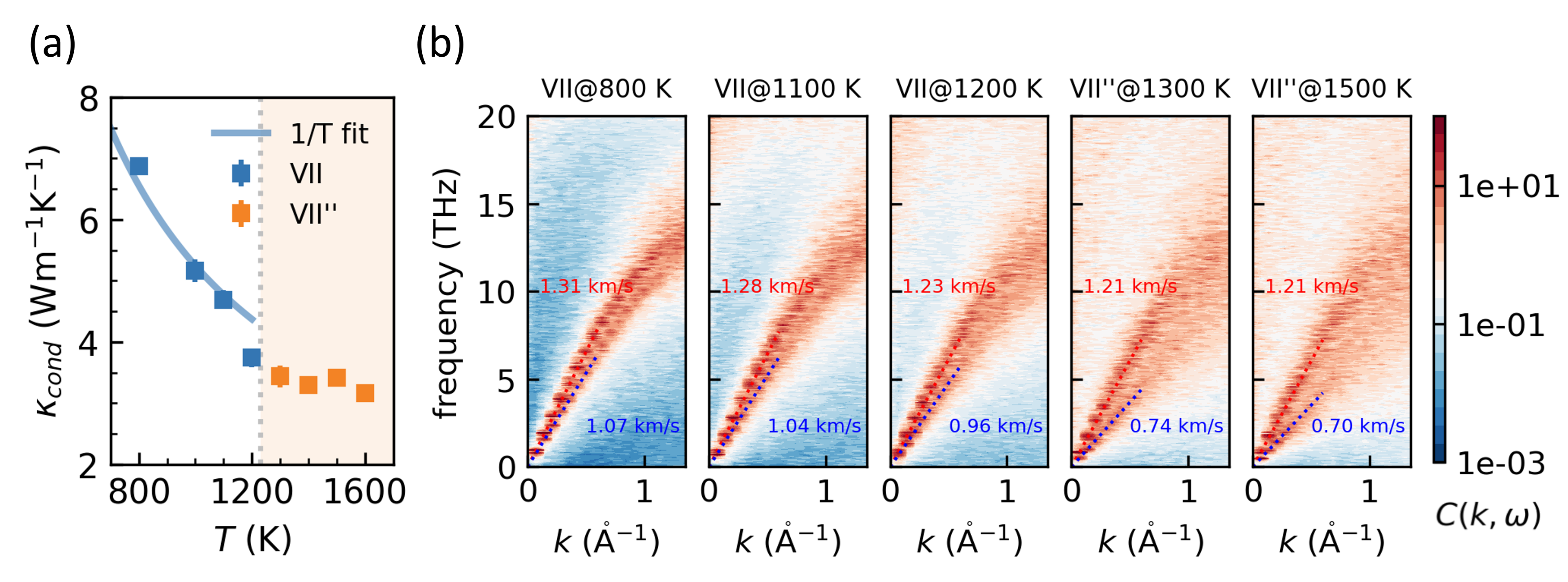}
  \caption{Temperature dependence of (a) thermal conductivity contributed by heat conduction $\kappa_{cond}$ and (b) normalized spectra energy density $C(k,\omega)$, where the direction of wavevector $\mathbf k = (n_x \Delta k_x,0,0)$ is set to $x$ direction with the wavenumber resolution of $\Delta k_x = 2 \pi /L_x \sim 0.07\ \rm{\mathring A^{-1}}$. The red and blue dotted line denotes the linear dispersion relationship for longitudinal and transverse acoustic branches, respectively.}
  \label{fig:4}
\end{figure*}

\textit{Transverse mode softening across superionic transition.} We also extract the contributions of heat conduction to $\kappa$ ($\kappa_{cond}$). As presented in Fig. \ref{fig:4}(a), $\kappa_{cond}$ shows a monotonic decrease trend for ice-VII. The $1/T$ dependence of $\kappa_{cond}$ is a typical characteristic that reveals the dominant role of three-phonon scatterings \cite{ref:roufosse1973}. At low temperatures, $\kappa_{cond}$ is much larger than $\kappa_{conv}$, leading to a decreasing trend. Therefore, the anomalous non-monotonic trend of $\kappa$ originates from the competing mechanism of heat conduction and convection. 

We also note a discontinuity of $\kappa_{cond}$ near the phase boundary. On one hand, it can be attributed to the density decrease accompanied by the VII-VII" phase transition (see Fig. S5 in SI).
On the other hand, a softening of the transverse acoustic modes is observed. Here we calculate the spectra energy density $C(k,\omega)$ from DPMD trajectories (see details in SI). The $C(\mathbf k,\omega)$ provides information on collective vibrational modes (group velocity, lifetime) inside the complex ice polymorph. As shown in Fig.\ref{fig:4}(b), we present the $C(k,\omega)$ in the low-frequency regime ($\nu \le 20\ \rm{THz}$), which contains the longitudinal and transverse acoustic branches that make dominant contributions to $\kappa_{cond}$. As the wavenumber approaches the center of the first Brillouin zone, the dispersion relationship exhibits a linear behavior, and the sound velocity can be extracted. For ice-VII, the sound velocities of both longitudinal and transverse acoustic branches decrease slightly with increasing temperature. For ice-VII", the sound velocity of longitudinal acoustic branches maintains a slight decreasing trend while the sound velocity of transverse acoustic branches shows a sudden larger decrease across the superionic transition. A significant softening of the transverse acoustic modes across the superionic transition is observed with decreased group velocity from $0.96\ \rm{km/s}$ to $0.74\ \rm{km/s}$. Moreover, as temperature increases, the vibrational energy peaks of both longitudinal and transverse acoustic branches exhibit broadening, corresponding to the reduction in phonon lifetimes.

Overall, by summarizing our results, we can understand the anomalous behavior of $\kappa$ in ice-VII across the superionic transition. At moderate temperatures, the propagation of lattice vibration mode dominates the $\kappa$ and exhibits a typical $1/T$ dependence due to the three-phonon scattering process.
At T = 1000 K, the onset of hydrogen diffusion between O-H···O pairs leads to an exponential increase by two orders of magnitude in the diffusion coefficient of protons. These protons hop within the oxygen-formed BCC sub-lattice and carry heat, creating a non-negligible contribution via heat convection.
As temperature increases above the superionic transition threshold (T = 1250 K), the first-order phase transition occurs. The significantly enhanced mobility of hydrogen, combined with softening in transverse acoustic branches, leads to a saturated value above T = 1300 K. Under such conditions, the diffusing protons can experience a stronger thermal diffusion process as compared with oxygen atoms.

In conclusion, by utilizing the newly developed DP-SCAN water model, we investigate the microscopic mechanism behind the anomalous thermal and proton transport of high-pressure ice across the superionic transition. We explain the anomalous increasing trend of $\kappa$ with elevated temperature and illustrate the important role of proton diffusion in superionic ice. To overcome the limitation of the traditional lattice dynamics approach which requires high-order force constants to correct the quasi-harmonic approximation, here we extract all the mass transport, heat transport, and collective dynamics from long-time large-scale molecular dynamics trajectories, without any assumptions or approximations made. These approaches combined with \textit{ab initio} accurate deep neural network potential energy surface model, can be applied to various complex materials, including proton-disorder ice polymorphs, proton-diffusion superionic crystal, and amorphous materials.

\section*{Acknowledgment}
This work was supported by the National Key R\&D Program of China under Grant No. 2017YFA0403200, the NSAF under Grant No. U1830206, the Science and Technology Innovation Program of Hunan Province under Grant No. 2021RC4026.

\bibliography{reference}

\end{document}